# Angular-asymmetric transmitting metasurface and splitter for acoustic wave: a combining of the Coherent Perfect Absorber and Laser

Shuting Cao, Zhilin Hou*

School of Physics and Optoelectronics, South China University of Technology, Guangzhou

510640, China

*phzlhou@scut.edu.cn

Abstract

Coherent Perfect Absorber (CPA) and Laser is a pair of inversely worked wave structure satisfying the time-reversal relation. We point out that, if the wave energy absorbed by CPA can be recovered by somewhat way and then be reemitted by its corresponded Laser, the combination of CPA and Laser will be a wave-controlling device that can "move" the wave energy from one place to another. By extending the concepts of CPA and Laser from the previously studied one-dimensional non-Hermitian systems to the two-dimensional ones, this understanding is used to simplify the designing of the metasurface in this letter. As examples, an angular-asymmetric transmitting metasurface and wave splitter, which acts as perfect transmitter/retro-reflector for waves coming from two oppositely tilted angles and split an incident wave into two directions with arbitrary amplitude ratio and phase difference, respectively, are constructed by combining two pieces of the separately designed CPAs. The idea can not only greatly simplify the designing of metasurface, but also bridge the researches between the metasurface and non-Hermitian system.

Coherent perfect absorber (CPA) and Laser is a special pair of wave device that satisfying the time-reversal symmetry. The relationship between them tells that, by replacing the loss (gain) materials by the gain (loss) ones in CPA (Laser), the perfect absorption (lasing) procedure can be reverted exactly as the emission (absorption) one[1-10]. This relationship and the behind physics have been suggested to be used in the designing of absorptive interferometer[3, 4], perfect wave absorbers in subwavelength[5, 6], light-surface plasmas coupler[9], lens without aberration[11] and others. However, those functional structures are based mostly on the coherent perfect absorption effect, which means only the procedure of removing the wave energy away from space has been used, while the anti-effect, say, the lasing-effect, which can bring the wave energy into space has in fact seldom been used in the structure designing. We point out that, if the wave energy absorbed by the CPA can be recovered by somewhat way and then be reemitted by its corresponded lasing structure, a combination of the CPA and Laser will act as a special device that can "move" the wave energy from one place to another. This will give a new mechanism for wave-behavior-controlling device.

On the other hand, controlling the wave behavior by artificial structures has attracted a lot of attention in recent years. The concept of metasurface, namely a two-dimensional (2D) thin artificial structure, have been introduced firstly in electromagnetic wave system[12] and extended into acoustic one[13]. They provide unique functionalities with large potential of engineering applications such as anomalous refraction and reflection[14-18], asymmetric transmission[19, 20], perfect absorptions[21, 22], retro-reflection[23], cloaking[24] and others[25, 26]. It can be found however that, the devices to realize those functionalities are mostly constructed by the phase-gradient approach[17] or its improved approaches[18], which often provides very complicated building subunits. We notice that, the designing purpose of the metasurface is very similar to the above mentioned CPA-Laser pair if we view it as the structure that can move the wave energy in real space from one part to another, or say, from one channel to the other. This similarity and the simple corresponding relationship between the CPA and Laser give us a new mechanism for the metasurface designing.

As a direct application, we will show that the above mentioned mechanism can be used to design the transmission-type metasurface. As we know, a transmission-type metasurface is a device designed to transmit an incident wave from the incident side to the transmission side in a designable direction. It is equivalent to a CPA-Laser pair because it "absorbs" the wave from the inputting side and "emits" it into the outputting side. This means that the inputting and outputting ends of the metasurface can be separately designed as a CPA and Laser. We will see that such a separately-designing strategy can greatly simplify the designing procedure. In this letter, we will take two examples to show the design procedure. As the first example, we will show that an angular asymmetric transmission metasurface (AATM) can be constructed by combining two pieces of the angular asymmetric reflective metausurface (AARM). As has been reported in Ref.[27] for electromagnetic wave and in Ref.[28] for acoustic wave, an AARM is a specially designed CPA that can perfectly absorb the waves from positive tilted incident direction and can in the same time perfectly retro-reflect the waves from negative tilted incident direction. We show that, by combining two pieces of AARM into a new structure, the combination will work in CPA status at the incident end and in Laser status at the outputting end. As a result, an AATM, which acts on both sides as perfect transmitter/retro-reflector for waves coming from two oppositely tilted angles, can be obtained. As the second example, we will show that the mechanism can also be used to design a transmitting wave splitter: a more general metasurface that can split the plane wave from $+\alpha$ direction in the inputting side into two plane waves in $\pm\alpha$ directions in the outputting side with arbitrary amplitude ratio and phase difference.

We start the designing from the AARM. The two-dimensional (2D) structure is shown schematically in Fig. 1(a), which is a structured rigid periodic surface with $L$ ($L$=2 is shown in the figure) bottomed grooves and one bottomless channel per period. By neglecting the wave dissipation in the medium, the bottomed grooves will act as the lossless material with pure imaginary surface impedance at their opening end, while for the bottomless channel, because the reflection from the outgoing end (not shown in the figure) will be smaller than unity, it will act as the loss material with complex surface impedance at the inputting end. The purpose of the designing is to find the configuration of the grooves and the channel, so that the surface can act as a CPA when it is illuminated by the incident waves from one or two given directions.

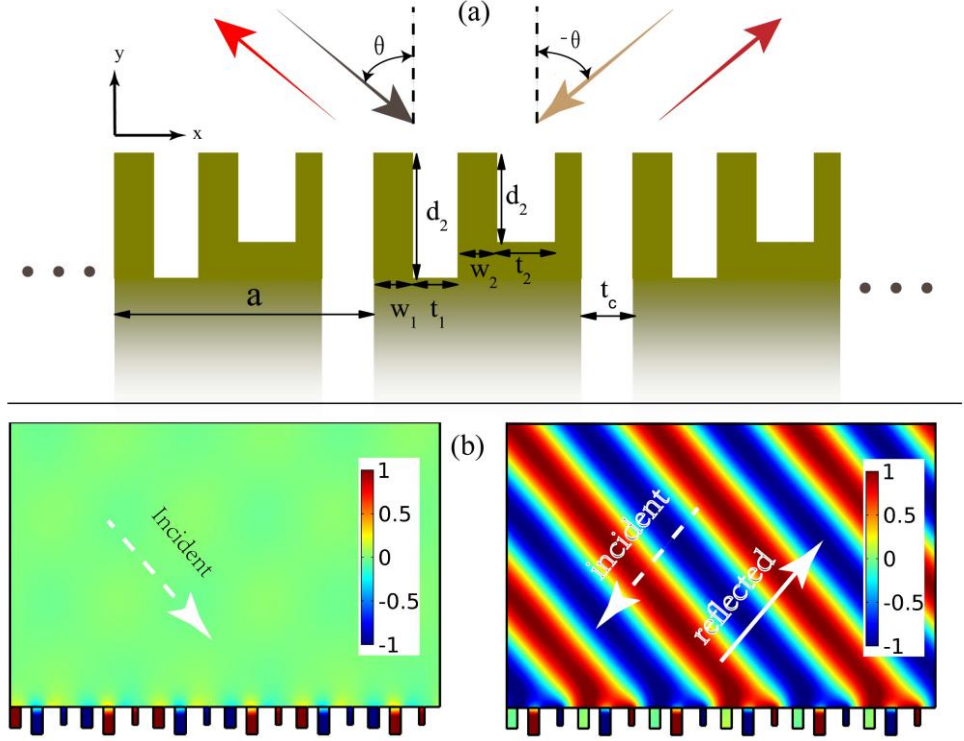

Fig. 1. (a) Schematic illustration of the designed AARM, which is a 2D planar period sound hard surface with bottomed grooves and bottomless channel. The periodicity is along *x*-direction, and the period is *a*. One (or two) incident wave from $\theta$ (or $\pm\theta$) direction(s) can be scattered as the diffractive modes. The depth and width of the grooves in each period are denoted as $t_l$, $d_l$ ($l=1,2,...,L$), the width of channel is denoted as $t_c$, and the relative distance between the *l*th and (*l+1*)th grooves is denoted as $w_l$, respectively. Because reflection from the outgoing end (not shown in the figure) is permitted, the channel will act as the loss material with complex surface impedance at the imputing end. (b) The scattered pressure field distributions for the designed AARM. In which the left and right panel shows respectively the result by the incident plane wave from +50° and -50° direction. The direction of the plane waves are shown by white arrows. Notice that the channels [the $(3l+1)^{th}$, ($l=0,1,\cdots$) groove in each panel] in the figure are replaced by grooves with the depth and the impedance calculated by $d_c = \phi_{1c}/2k_0$ and $Z_c=(1+r_{1c})/(1-r_{1c})$ respectively.

Rather than using the phase-graded method suggested in the Ref.[28], we introduce here a new method based on the general grating theory to design the structure. The new method can give simple structure and clear understanding of the behind physics. Because of the *x*-directional periodicity of the structure, the pressure and the y-component particle velocity of the field in the medium above the surface can be written as a summation of the harmonic modes:

$$p = \sum_n A_n^+ e^{-j(k_0 \sin\theta + G_n)x} e^{jk_0\beta_n y} + \sum_n A_n^- e^{-j(k_0 \sin\theta + G_n)x} e^{-jk_0\beta_n y}$$

$$v = -\frac{1}{Z_0}\sum_n \beta_n A_n^+ e^{-j(k_0\sin\theta+G_n)x}e^{jk_0\beta_n y} + \frac{1}{Z_0}\sum_n \beta_n A_n^- e^{-j(k_0\sin\theta+G_n)x}e^{-jk_0\beta_n y} \quad (1)$$

where $k_0 = 2\pi/\lambda_0$ is the wave vector with $\lambda_0$ as the working wavelength, $\theta$ is the incident angle between the incident ray and the surface normal, it takes positive (negative) value if the incident ray is at the left- (right-) hand side of the surface normal; $Z_0$ is the characteristic impedance of the medium, $A_n^+$ and $A_n^-$ is the amplitude of the $n^{th}$-order incident and diffractive harmonic mode, respectively; $G_n = \dfrac{2n\pi}{a}; n=0,\pm 1,\cdots$ is the Bloch wave vector with $a$ as the period of the structure, and $\beta_n = \pm\sqrt{1-(\sin\theta+n\lambda_0/a)^2}$ denotes the normalized y-directional propagation constant of the $n^{th}$-order harmonic mode. It can be seen from the formula that $\beta_n$ can be real, i.e., the $n^{th}$-order diffractive mode can be propagative only when the condition $|\sin\theta + n\lambda_0/a| < 1$ is satisfied. By setting $a = \lambda_0/2\sin\alpha$, the system will have only the $0^{th}$- and $-1^{st}$-order harmonic modes as the propagating ones when $\theta$ satisfying the condition $2\sin\alpha - 1 < \sin\theta < 4\sin\alpha - 1$. As we will show later that, the system is equivalent to a two-port non-Hermitian system under this setting, and the design target is to find the spectrum singularity of the system, at which the incident plane waves with incident angle $\theta = \pm\alpha$ can be perfectly "absorbed".

On the other hand, the pressure and y-component particle velocity field in grooves and channel can be expressed as the superposition of the waveguide modes, the field at the inputting end (y=0) of the $l^{th}$ groove (or channel) can be expressed generally as

$$p_l = \sum_k H_{kl}^+ \cos\frac{(k-1)\pi}{t_l}(x-x_l)\left(1+\left|\frac{H_{kl}^-}{H_{kl}^+}\right|e^{-j\phi_{kl}}\right)$$

$$v_l = -\frac{1}{Z_0}\sum_k H_{kl}^+ \sigma_{kl} \cos\frac{(k-1)\pi}{t_l}(x-x_l)\left(1-\left|\frac{H_{kl}^-}{H_{kl}^+}\right|e^{-j\phi_{kl}}\right) \quad (2)$$

where $x_l$ and $t_l$ is the position and width of the $l^{th}$ groove (or channel), respectively, $H_{kl}^+$ ($H_{kl}^-$), ($k=1,2,\ldots$) is the amplitude of the $k^{th}$-order waveguide mode propagating into (out from) the $l^{th}$ groove (or channel), $\phi_{kl}$ is the phase difference between $H_{kl}^+$ and $H_{kl}^-$. By defining the normalized propagating constant of the $m^{th}$-order waveguide mode in the $l^{th}$ groove (or channel) as $\sigma_{kl} = \sqrt{1 - \left(\frac{(k-1)\lambda_0}{2t_l}\right)^2}$, $k = 1, 2, \cdots$, and defining the reflection ratio as $r_{kl} = \left|\frac{H_{kl}^-}{H_{kl}^+}\right|$, we have always $r_{kl} = 1$ and $\phi_{kl} = k_0\sigma_{kl}2d_l$ for bottomed grooves. While for bottomless channel, we can introduce energy loss by restricting $r_{kc} < 1$, here and in the follows, we let $l=c$ in the subscripts to denote the terms in channel.

By Eqs.(1), (2) and the continuum boundary condition at the surface, we can obtain a linear equation set for variables $A_n^-$ and $H_{kl}^+$ under given $A_n^+$, $a$, $x_l$, $t_l$, $d_l$, $x_c$, $t_c$, $r_{kc}$ and $\phi_{kc}$. With the linear equation set, we can search the CPA structure by an optimization procedure. The detailed deduction of the linear equation set is presented in Note 1 of the supplementary material. In all of our optimization procedure (see Note 2 in supplementary material for details), we set the width of the channel $t_c$ to be $0.15a$, so that only the $0^{th}$-order waveguide mode ($k=1$) can propagate in it, which means except $r_{1c}$ and $\phi_{1c}$, all other $r_{kc}$ and $\phi_{kc}$ ($k=2, 3, \cdots$) will be set as zero. The position of the channel is set as $x_c = 0$, and the restriction on depths of grooves are chosen as small as possible (restricted in the region $d_l < 0.3\lambda_0$ in calculation). To avoid structure with extremely thin wall, narrow groove or narrow channel, $w_l$ and $t_l$ are restricted to be greater than $0.05a$.

As the first example, we use this scheme to search for the AARM, or say, the unidirectional CPA that can absorb the plane wave from a single direction with incident angle $\theta = \alpha$. Notice that for a reciprocal two-port system (i.e., metasurface with only $0^{th}$- and $-1^{st}$-order possible diffractive modes), the perfect absorption of the wave from $\theta$ direction means also a retro-reflection for the wave from $-\theta$ direction. By the method, structures with $\alpha$ from 30° to 88° are searched and found. We find that for all of these structures, 2 grooves per period are enough to realize the function. As examples, we choose the structure with $\alpha = 50°$ to show the result. The optimized parameters are obtained as $d_{1,2} = (0.249, 0.163)\lambda_0$, $w_{1,2} = (0.152, 0.256)a$, $t_{1,2} = (0.153, 0.076)a$, $r_{1c} = 0.634$ and $\phi_{1c} = 2.403$. To verify the effect, a finite element simulations based on Comsol Multiphysics is performed (see Note 3 in Supplementary material for details). The scattered pressure field for incident plane waves from $\theta = +50°$ and $-50°$ directions are presented in the left and right panels in Fig. 1(b), respectively. In the simulation, the channels [the $(3l+1)^{th}$, ($l=0,1,\cdots$) groove from left in each panel] are replaced by the bottomed grooves with depth $d_c = \phi_{1c}/2k_0 = 0.191\lambda_0$ and with the impedance $Z_c = Z_0(1+r_{1c})/(1-r_{1c}) = 4.469Z_0$ at the bottom. It can be found from the figure that the maximum amplitude of the reflective wave in the left panel is about 0.025 and in the right panel is about 0.999 under the incident amplitude of unity. This means the effect of absorbing (retro-reflecting) the incident wave from positive (negative) 50° direction is perfect.

In this obtained unidirectional CPA, if we turn around the directions of the waveguide modes in the channel (i.e., $H_{1c}^-$ is now the incident wave and $H_{1c}^+$ is the reflective wave) and then define the reflection ratio as $r_{1c}^{'} = \frac{|H_{1c}^+|}{|H_{1c}^-|} = \frac{1}{r_{1c}}$, we will have the impedance at the bottom of the channel as $Z_c^{'} = Z_0 \frac{1+r_{1c}^{'}}{1-r_{1c}^{'}} = -Z_0 \frac{1+r_{1c}}{1-r_{1c}} = -Z_c$, which means the structure turns from loss to gain. This result means that the CPA and its corresponded time-reversal structure, i.e., the Laser, are in fact the same structure: it works in CPA status when it is illuminated by the wave from outside of the structure, while works in the Laser status when it is illuminated by the waveguide mode inside the channel. With this understanding, we can construct the angular asymmetric refractive metausurface simply by sticking together the CPA and Laser substructure back-to-back with their channels connected. Notice that because there is only one channel per period, we can have two different ways to stick the substructures: to stick them directly back-to-back (refers as D-type in the follows) or to rotate first one of them by a π angle along y axes and then stick them back-to-back (refers as R-type in the follows). Notice also that, because the depth of the channel calculated by $d_c = \phi_{1c}/2k_0$ is usually smaller than the maximum value of the ones of the grooves, the total depth of the channel (and the thickness of the metasurface) should take the value as $d_T=2d_c+\lambda/2$ to avoid the overlapping of the grooves in thickness direction. As we know, according to the formula $Z = \frac{1+r_{1c}e^{-j2k_0 d_T}}{1-r_{1c}e^{-j2k_0 d_T}}$, a change of $n\lambda/2$ ($n$ is an integer) of $d_T$ would not change the value of the impedance at the opening of the channel. Shown in Fig. 2 is the scattered pressure field for the system with $\alpha=50°$ under the incident wave with $\theta=\alpha$, in which the results for the D-type and R-type structure are shown respectively in the left and right part. The thickness of the whole structure is obtained as $d_T=0.882\lambda_0$. It can be seen that the transmission angle for D-type and R-type structure is -50° and +50°, respectively, and the amplitude of the transmitted pressure wave is obtained as 0.999 under the incident amplitude of unity for both structures, which means the almost perfect transmission is obtained. We do not show the situations for incident wave from θ=-50 ° because it is the same as the one shown the right panel of Fig.1(b). And because of the reciprocity of the structure, the situation will be the same when the incident wave is from the lower half-space of the metasurface. Those results show that an AATM, which acts on both sides as perfect transmitter/retro-reflector for waves coming from two oppositely tilted angles, is obtained.

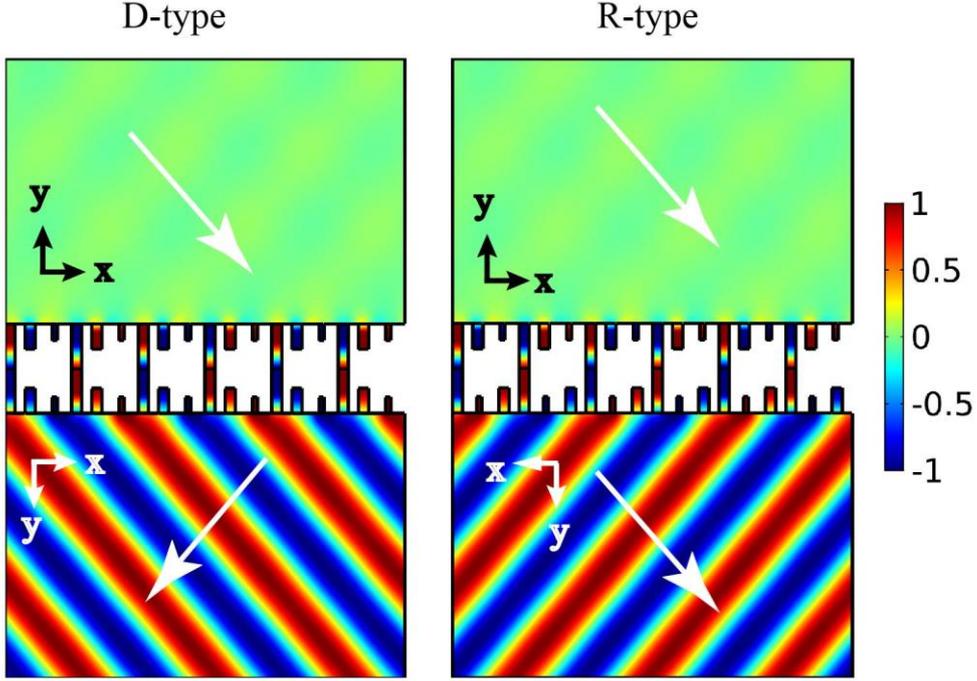

Fig.2 The scattered pressure field distribution of the designed AATM illuminated by plane wave from +50 ° direction. The left panel is for the D-type structure, and the right panel is for the R-type structure. The white arrows are given to show the directions of the incident and transmitted plane waves. For clear eyesight, the incident fields are not shown in the figure. The coordinate systems in each panel are given to show how the AATMs are combined by rotating and sticking the AARMs.

When we focus only on the scattering behavior of the waves in the upper or lower half-infinite space of the AATM, the structures can also be described as a two-port non-Hermitian system: a topic that has been attracting a lot of interest in recent years[8, 29, 30]. One may find that, previous works on this topic concentrated mostly on the one-dimensional (1D) system. Here we show that our 2D case has the same property as of the 1D one. Such an extension will bridge the researches between the metasurface and the non-Hermitian system.

As is schematically shown as inset in Fig. 3(a), by expressing respectively the incoming and outgoing waves from the left- (right-) hand side of the surface-normal directions as $p_L^+$ ($p_R^-$) and $p_L^-$ ($p_R^+$), the scattering property of the metasurface in the upper (or lower) half-infinite space can be described as the scattering matrix as

$$\begin{pmatrix} p_R^+ \\ p_L^- \end{pmatrix} = \begin{pmatrix} t_P & r_N \\ r_P & t_N \end{pmatrix} \begin{pmatrix} p_L^+ \\ p_R^- \end{pmatrix}. \qquad (3)$$

We can find that the scattering matrix presented in the equation has only one nonzero element (with $|r_N|=1$) at the status of angular-asymmetric-reflecting. This status is called also as the spectrum singularity in a non-Hermitian system[1, 10]. A checking of the spectra $|r_P|$, $|t_P|$ (also $|t_N|$), $|r_N|$, and their phases $\varphi_{r_P}$, $\varphi_{t_P}$ and $\varphi_{r_N}$ as functions of the wavelength $\lambda$ of the incident wave are shown in Fig. 3(a) and (b), in which the singularity at $\lambda_0$ can be clearly seen. Specially, the phase diagram $\varphi_{r_P}$ and $\varphi_{t_P}$ undergoes an abrupt $\pi$ phase shift at $\lambda_0$. This implies the existence of the divergence of their derivative with respect to the wavelength resulting in the diverging delay time of the reflected or transmitted wave[6]. This huge delay time increases the interaction of the waves with the "loss" element, causes the so-called coherent perfect absorption behavior.

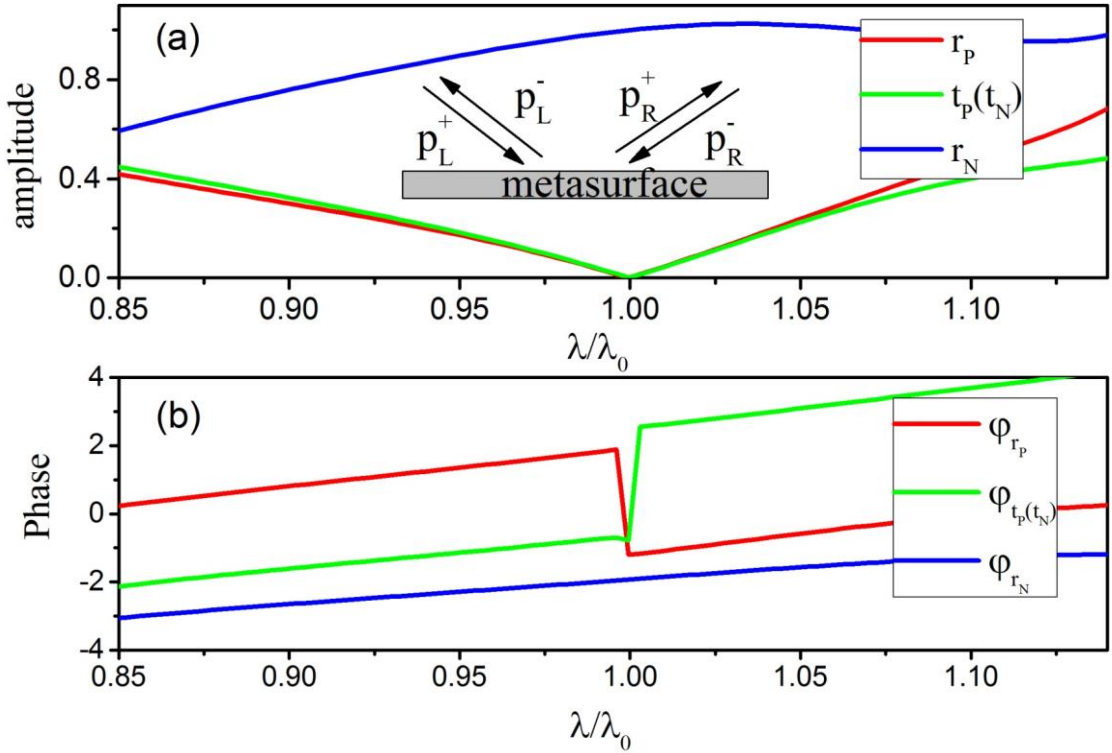

Fig.3 (a) amplitudes and (b) phases of the elements in the scattering matrix as the function of the wavelength of the incident wave. The inset in (a) is given to show that the scattering behavior of the waves in the upper half-infinite space of the AATM can be described as a two-port non-Hermitian system.

The above mentioned unidirectional CPA can be extended as the general case that can perfectly absorb two plane waves from $\pm\alpha$ directions with arbitrary amplitude ratio and phase difference. By such an extension, more complicated functional metasurfaces can be realized. As our second example, we show that a wave splitter, that can split a plane wave from α direction into two in-phase plane waves with equal amplitude in $\pm\alpha$ directions, can be designed. To construct such a structure, we need a unidirectional CPA as the inputting end, and in the same time a general CPA that can absorb perfectly two plane waves from $\pm\alpha$ directions with the same amplitude and phase as the outputting end. Shown in Fig. 4 is the scattered pressure field for the structure with α=50°, in which the wave splitting phenomenon can be clearly seen. In the figure, we use a Gaussian beam as the incident wave to show intuitively the splitting effect. It is necessary to point out that to keep the impedance at the opening of the channel unchanged before and after the combination, the $r_{1c}$ value for the inputting and outputting end should be the same one. In practice, we can first search the structural parameters $w_l^I, t_l^I, d_l^I, r_{1c}$ and $\phi_{1c}^I$ for given $t_c$ for the inputting end, with the obtained $r_{1c}$ and the given $t_c$, we then search the structural parameters $w_l^O, t_l^O, d_l^O$ and $\phi_{1c}^O$ for the outputting end, here the superscript "I" and "O" means respectively the parameters for the inputting and outputting end. By this method, the thickness of the combined structure can be obtained as $d_T = d_c^I + d_c^O + \lambda/2$, where $d_c^{I(O)}$ is calculated by the formula $\phi_{1c}^{I(O)}/2k_0$. For the structure shown in Fig.4, we use the same structure given in Fig. 2(b) as the inputting end. Under its $r_{0c}$ and $t_c$, the structural parameters for the outputting end are obtained as $d_{1,2}^O = (0.188, 0.245)\lambda_0$, $w_{1,2}^O = (0.156, 0.063)a$, $t_{1,2}^O = (0.373, 0.121)a$ and $\phi_{1c}^O = 2.00$, and the total thickness of the structure is obtained as $d_T$=0.851$\lambda_0$.

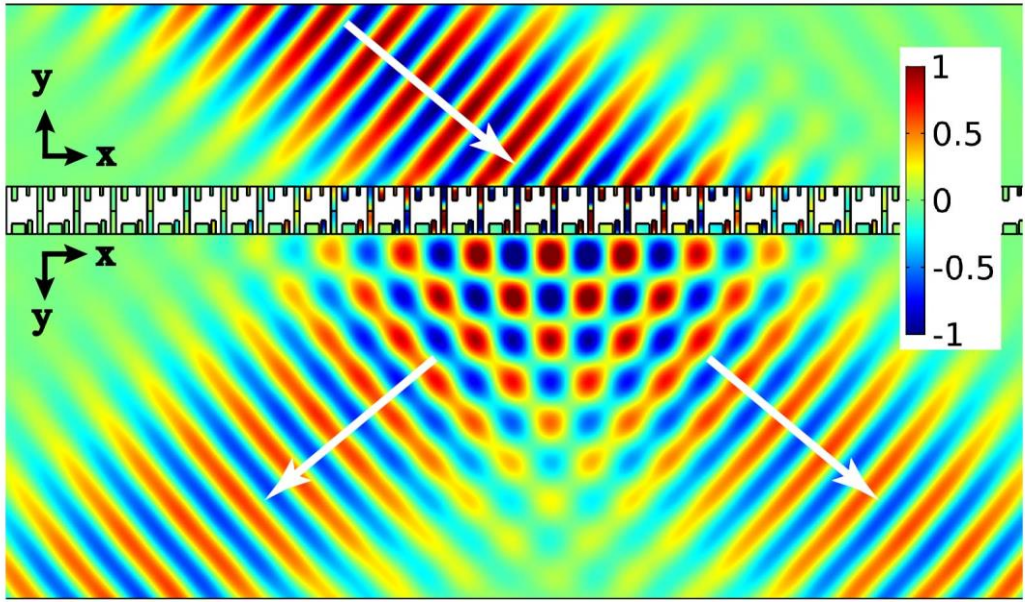

Fig.4 Pressure field distribution of the wave splitter designed to split the wave from +50 ° direction into two in-phase waves with equal amplitude in ±50 ° directions. The white arrows show the incident and transmitted beams.

In conclusion, we show that an AATM and a wave splitter can be constructed by combining a CPA and its corresponded Laser. Because the inputting and outputting ends are decoupled and separately designed, the suggested mechanism gives a simple designing method for transmission-type of metasurface. The scattering property of the structure in the inputting (or outputting) half-infinite space is also studied by the scattering matrix method and found to be as the same as the one for the one-dimensional non-Hermitian system. The research not only gives a simple way for the metasurface designing, but also bridges the researches between the metasurface and the non-Hermitian system.